\documentclass[aps,prx,reprint,longbibliography,superscriptaddress]{revtex4-2}

\usepackage{amsmath,amssymb,bm}
\usepackage{graphicx}
\usepackage{hyperref}
\usepackage{xcolor}

\hypersetup{
  colorlinks=true,
  linkcolor=blue,
  citecolor=blue,
  urlcolor=blue
}

\newcommand{\sro}{Sr$_2$RuO$_4$}
\newcommand{\musr}{$\mu$SR}
\newcommand{\tc}{\ensuremath{T_{\rm c}}}
\newcommand{\tlambda}{\ensuremath{T_{\Lambda}}}
\newcommand{\ttrsb}{\ensuremath{T_{\rm TRSB}}}
\newcommand{\hcTwo}{\ensuremath{H_{\rm c2}}}

\begin{document}

\title{Normal-state exponential muon-spin relaxation and its implications for time-reversal symmetry breaking in Sr$_2$RuO$_4$}

\author{Rustem Khasanov}
 \email{rustem.khasanov@psi.ch}
 \affiliation{PSI Center for Neutron and Muon Sciences, 5232 Villigen PSI, Switzerland}

\author{Thomas J. Hicken}
 \affiliation{PSI Center for Neutron and Muon Sciences, 5232 Villigen PSI, Switzerland}

\author{Jonas A. Krieger}
 \affiliation{PSI Center for Neutron and Muon Sciences, 5232 Villigen PSI, Switzerland}

\author{Maxime Lamotte}
 \affiliation{PSI Center for Neutron and Muon Sciences, 5232 Villigen PSI, Switzerland}

\author{Toni Shiroka}
 \affiliation{PSI Center for Neutron and Muon Sciences, 5232 Villigen PSI, Switzerland}

\author{Jake S. Bobowski}
 \affiliation{Department of Physics, University of British Columbia, Kelowna, BC V1V 1V7, Canada}
 \affiliation{Toyota Riken - Kyoto-Univ. Research Center (TRiKUC), Kyoto 606-8501, Japan}

\author{Vadim Grinenko}
 \affiliation{Tsung-Dao Lee Institute \& School of Physics and Astronomy, Shanghai Jiao Tong University, Shanghai 201210, China}

\author{Thomas Johnson}
 \affiliation{Department of Engineering, Kyoto University of Advanced Science (KUAS), Kyoto 615-8577, Japan}
 \affiliation{Toyota Riken - Kyoto-Univ. Research Center (TRiKUC), Kyoto 606-8501, Japan}

\author{Naoki Kikugawa}
 \affiliation{National Institute for Materials Science, Tsukuba 305-0003, Japan}

\author{Hans-Henning Klauss}
 \affiliation{Institut f\"{u}r Festk\"{o}rper und Materialphysik, Technische Universit\"{a}t Dresden, 01062 Dresden, Germany}

\author{Shubhankar Paul}
\affiliation{Department of Physics, Indian Institute of Technology Kanpur, Kanpur 208016, India}
\affiliation{Toyota Riken - Kyoto-Univ. Research Center (TRiKUC), Kyoto 606-8501, Japan}

\author{Benjamin M. Huddart}
\affiliation{Clarendon Laboratory, University of Oxford, Department of Physics, Oxford OX1 3PU, United Kingdom}

\author{Stephen J. Blundell}
\affiliation{Clarendon Laboratory, University of Oxford, Department of Physics, Oxford OX1 3PU, United Kingdom}

\author{Hubertus Luetkens}
 \affiliation{PSI Center for Neutron and Muon Sciences, 5232 Villigen PSI, Switzerland}

\author{Yoshiteru Maeno}
 \email{maeno.yoshiteru.b04@kyoto-u.jp}
 \affiliation{Department of Engineering, Kyoto University of Advanced Science (KUAS), Kyoto 615-8577, Japan}
 \affiliation{Toyota Riken - Kyoto-Univ. Research Center (TRiKUC), Kyoto 606-8501, Japan}
 \affiliation{Max Planck Institute for Chemical Physics of Solids (MPI-CPfS), N\"{o}thnitzer Str. 40, 01187 Dresden, Germany}

\date{\today}

\begin{abstract}
%
\noindent The low-temperature increase of the zero-field muon-spin relaxation rate in \sro\ has long been regarded as evidence for spontaneous internal fields associated with time-reversal-symmetry-breaking (TRSB) superconductivity. Here we directly test this interpretation using muon-spin rotation/relaxation (\musr) measurements on high-quality \sro\ single crystals. For all four samples, the onset temperature of the additional exponential relaxation in the zero-field \musr\ data, \tlambda, lies below the bulk superconducting transition temperature \tc, even in the absence of applied strain. More importantly, in transverse-field \musr\ measurements we find a  quantitatively similar exponential relaxation in a field of 80~mT applied along the crystallographic $c$ axis. This field exceeds the upper critical field of the crystals studied in transverse-field geometry and suppresses bulk superconductivity, demonstrating that the characteristic relaxation persists in the normal state. Measurements below the upper critical field reveal an additional superconducting mixed-state contribution superposed on this normal-state channel. Our results show that the zero-field \musr\ relaxation in \sro\ cannot, by itself, be regarded as a unique signature of superconducting TRSB. They call for a reassessment of a key experimental signature of TRSB in \sro\ and highlight the need to distinguish superconducting and normal-state relaxation channels in other unconventional superconductors.
\end{abstract}

\maketitle
\section{Introduction}

\sro, despite its modest superconducting transition temperature $T_{\rm c}$
of about 1.5~K, remains a central material in the study of unconventional
superconductivity. Its normal-state electronic structure is exceptionally
well characterized, but the superconducting order parameter is still debated
three decades after the discovery of superconductivity
\cite{Maeno1994,Mackenzie2003,Maeno2024,Maeno2024NatPhys}.
Early experiments were interpreted as supporting a spin-triplet chiral
$p$-wave state. In particular, early $^{17}$O nuclear magnetic resonance
(NMR) Knight-shift measurements suggested an unchanged spin susceptibility
below \tc\ \cite{Ishida1998}, while zero-field muon-spin rotation/relaxation
(ZF-\musr) experiments detected an enhanced relaxation below \tc,
interpreted as the appearance of spontaneous internal fields due to
time-reversal symmetry breaking (TRSB) \cite{Luke1998,Luke2000}. Polar Kerr
measurements also detected a finite Kerr angle below \tc\ \cite{Xia2006},
and Josephson experiments were interpreted in terms of a complex order
parameter \cite{Kidwingira2006,Anwar2013}.

The original chiral $p$-wave scenario has, however, become difficult to
reconcile with later experiments. Revised NMR Knight-shift measurements
showed a decrease of the spin susceptibility below \tc\
\cite{Pustogow2019,Ishida2020,Chronister2021}, and polarized neutron
scattering reached a compatible conclusion \cite{Petsch2020}. A recent
high-precision muon Knight-shift experiment also found a reduction of the
Knight shift below \tc, providing an independent \musr\ constraint
consistent with spin-singlet-like pairing \cite{Matsuki2026}. The present
debate has therefore shifted toward even-parity order parameters. If TRSB
is intrinsic to the superconducting state, possible candidates include
complex combinations such as $d_{xz}\pm i d_{yz}$, $s\pm i d$, or
$d\pm i g$ \cite{Grinenko2021NatCommun,Roemer2022}. These states generally
imply either a symmetry-protected two-component order parameter or an
accidental near-degeneracy of two one-dimensional representations
\cite{SigristUeda1991,Grinenko2021NatCommun,Roemer2022}. A distinct
possibility is inhomogeneous TRSB near defects or dislocations in an
otherwise time-reversal symmetric single-component superconducting state
\cite{Willa2021}.

Several recent thermodynamic and strain experiments have sharpened this
problem. Ultrasound experiments reported jumps in the shear elastic modulus
$c_{66}$ at \tc, which can be interpreted as evidence for a multicomponent
order parameter \cite{Ghosh2021,Benhabib2021}. For such a state, uniaxial
strain is expected to lift the degeneracy between the components and, in
general, to split the superconducting and TRSB transitions. A split between
\tc\ and \ttrsb\ was indeed reported by \musr\ under uniaxial stress
\cite{Grinenko2021NatPhys,Grinenko2023PRB}. However, this behavior has not
been confirmed by thermodynamic probes: high-sensitivity heat-capacity
measurements under uniaxial pressure found no evidence for a second
transition \cite{Li2021PNAS}. Elastocaloric measurements under uniaxial
strain further mapped the thermodynamic phase diagram of \sro, including
the superconducting phase boundary and a strain-induced magnetic phase
\cite{Li_Nature_2022}. More recently, high-resolution measurements of \tc\
and the elastocaloric effect under $\langle 110\rangle$ uniaxial stress
found no indication of the transition splitting expected for a homogeneous
two-component TRSB superconducting state \cite{Jerzembeck2024}. Thus, the
strain-induced split between \tc\ and the putative TRSB onset is presently a
$\mu$SR-specific observation. This discrepancy raises the question of whether
the onset of ZF-\musr\ relaxation indicates a bulk superconducting
transition. This question is particularly important for unstrained \sro,
where tetragonal symmetry is preserved and a homogeneous intrinsic TRSB
superconducting state would be expected to have $\ttrsb=\tc$.

\begin{figure}[htb]
\centering
\includegraphics[width=0.9\linewidth]{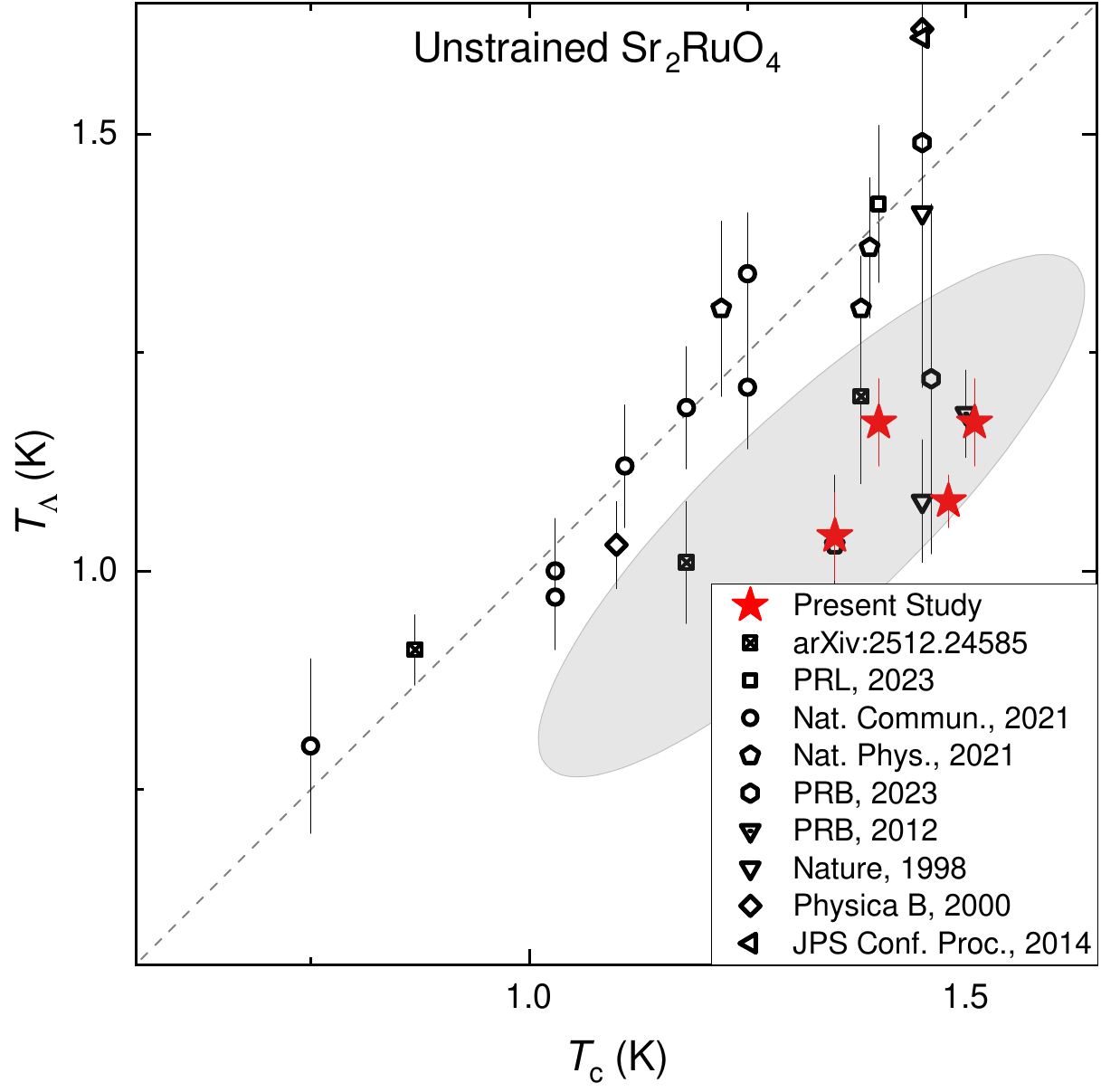}
\caption{\label{fig:literature}
Comparison between the bulk superconducting transition temperature \tc\ and
the onset temperature \tlambda\ of the exponential relaxation rate
$\Lambda$ for unstrained \sro. Red stars denote the present samples. Open and partly filled
symbols summarize data from previous ZF-\musr\ studies, including
Refs.~\cite{Luke1998,Luke2000,Shiroka2012,Higemoto2014,
Grinenko2021NatCommun,Grinenko2021NatPhys,Grinenko2023PRB,
Khasanov2023PRL,Li2025Preprint}. The dashed line corresponds to
$\tlambda=\tc$. The gray region highlights the samples showing the strongest
separation between \tc\ and \tlambda.}
\end{figure}

To address this question, we compare the available zero-field \musr\ results
on unstrained \sro\ from Refs.~\cite{Luke1998,Luke2000,Shiroka2012,
Higemoto2014,Grinenko2021NatCommun,Grinenko2021NatPhys,Grinenko2023PRB,
Khasanov2023PRL,Li2025Preprint}. The comparison is shown in
Fig.~\ref{fig:literature}, where the bulk superconducting transition
temperature \tc\ is plotted against the onset temperature $T_\Lambda$
extracted from the exponential ZF-\musr\ relaxation rate $\Lambda$. To place
the literature data on the same footing, the temperature dependence of the
exponential relaxation rate was fitted, where possible, with the
phenomenological form used in
Refs.~\cite{Grinenko2021NatPhys,Grinenko2021NatCommun}:
\begin{equation}
\Lambda(T)=
\begin{cases}
\Lambda_0, & T>\tlambda,\\
\Lambda_0+\Delta\Lambda\left[1-\left(T/\tlambda\right)^n\right],
& T<\tlambda,
\end{cases}
\label{eq:lambda}
\end{equation}
where $\Lambda_0$ is the high-temperature relaxation rate, $\Delta\Lambda$
is the low-temperature enhancement, and $n$ is a phenomenological power-law
exponent. Although many points lie close to $\tlambda=\tc$ (the dashed line
in Fig.~\ref{fig:literature}), a substantial fraction of the ZF-\musr\ data
show $\tlambda<\tc$. This spread indicates a nonuniversal relation between
the exponential relaxation and superconductivity. If \tlambda\ were the
onset of a homogeneous TRSB superconducting state, one would expect a robust
bulk transition and, when separated from \tc, a corresponding anomaly in
other probes.

In this paper we revisit the \musr\ evidence from a different perspective.
We study four unstrained \sro\ single crystals with different \tc\ values.
We compare zero-field (ZF) relaxation with transverse-field (TF) relaxation
measured at fields both below and above the upper critical field for
$H\parallel c$ ($\hcTwo^{\parallel c}$). Since the zero-temperature upper critical field
$\mu_0\hcTwo^{\parallel c}(0)$ is below 80~mT for all studied crystals, the
80~mT TF-\musr\ data provide an empirical measure of relaxation processes
present when bulk superconductivity is suppressed.
We also use muon stopping-site calculations and simulations
of the nuclear contribution to assess whether static nuclear moments can
account for the observed ZF relaxation.

We find that the low-temperature exponential relaxation observed in
ZF-\musr\ has a close counterpart in this normal-state data set. By
contrast, TF-\musr\ measurements below
$H_{\rm c2}^{\parallel c}(0)$ reveal an additional
relaxation associated with the vortex state, superposed on the same
underlying normal-state relaxation channel. This approach allows us to
separate superconductivity-related broadening from relaxation processes that
are already present in the normal state, and provides a direct test of
whether the ZF-\musr\ relaxation anomaly is a universal signature of bulk
TRSB in \sro.

\section{Experimental details}

The \musr\ experiments were performed on four \sro\ single crystals, denoted here as samples A--D. The corresponding original sample labels are A (C516E2R), B (C507BC), C (C501C), and D (C140). All samples were grown at Kyoto University \cite{Bobowski2019}. Details of the crystal-growth procedure, photographs of the crystals, and Laue x-ray images are provided in the Supplementary Information. The samples used in the present study are of high quality, as confirmed by their sharp superconducting transitions and by the small amount of impurity phases detected in dc-magnetization measurements performed at 10~mT and 1~T; see the Supplementary Information.

The superconducting properties of samples A--C were characterized by ac susceptibility, as described in the Supplementary Information. The superconducting transition temperatures were determined from the peak position of the imaginary part of the ac-susceptibility signal and were found to be \tc=1.51(1), 1.48(1), and 1.40(1)~K for samples A, B, and C, respectively. For sample D, the superconducting transition temperature was determined from specific-heat measurements and reported to be \tc=1.35(2)~K \cite{Grinenko2021NatCommun}.

The main set of zero-field (ZF), transverse-field (TF), and longitudinal-field (LF) \musr\ experiments was performed on samples A--C. For sample D, only ZF-\musr\ measurements were performed. Most of the \musr\ experiments were carried out on the FLAME \musr\ spectrometer at the Paul Scherrer Institute (PSI, Villigen, Switzerland), using a dilution refrigerator. These measurements were performed in ZF and in applied fields up to 1~T, over the temperature range from 40~mK to 2~K. The accessible temperature interval was set by the weak-thermal-link mode of operation, which allows the temperature to be scanned over this range without losing the $^3$He--$^4$He mixture. To exclude possible spectrometer- or cryostat-related effects, the ZF-\musr\ measurements on sample B were repeated on the VMS \musr\ spectrometer, also at PSI, using a $^3$He sorption-pump cryostat with a base temperature of 0.25~K. The data were analyzed using the MUSRFIT package \cite{MUSRFIT}.

The experimental geometry and data-analysis procedure are discussed in the Supplementary Information. The exponential relaxation rates obtained from fits of the ZF-, TF-, and LF-\musr\ data, which are the main focus of this paper, are denoted by $\Lambda^{\rm ZF}$, $\Lambda^{\rm TF}$, and $\Lambda^{\rm LF}$, respectively.

\textit{Use of AI tools.} Anthropic's Claude Opus 5 was used to write the analysis and plotting scripts that generate the tabulated coordinates, barriers, and zero-point energies, as well as Fig.~\ref{fig:muon-hopping} and Figs.~S13--S15 in the Supplementary Information; to assist with preparing, running, and analyzing the calculations; to help identify relevant literature; and to edit the text. Claude Fable 5.1 was used to help formulate the symmetry arguments used to identify candidate transition states. All arguments, calculations, and outputs were checked by the authors, who take full responsibility for the content.

\section{Results}

\subsection{Upper critical field}

Before comparing the ZF- and TF-\musr\ relaxation rates, it is necessary to establish the field-temperature boundary between the superconducting and normal states. The $T_{\rm c}(H)$ points for samples A--C were determined from zero-field-cooled (ZFC) and field-cooled (FC) TF-\musr\ measurements with $H\parallel c$. At each applied field, the transition temperature was identified from the onset of the additional Gaussian relaxation $\sigma^{\rm TF}$ and from the corresponding change in the internal field $B_{\rm int}$, as demonstrated for samples A--C in the Supplementary Information.

\begin{figure}[htb]
\centering
\includegraphics[width=0.9\linewidth]{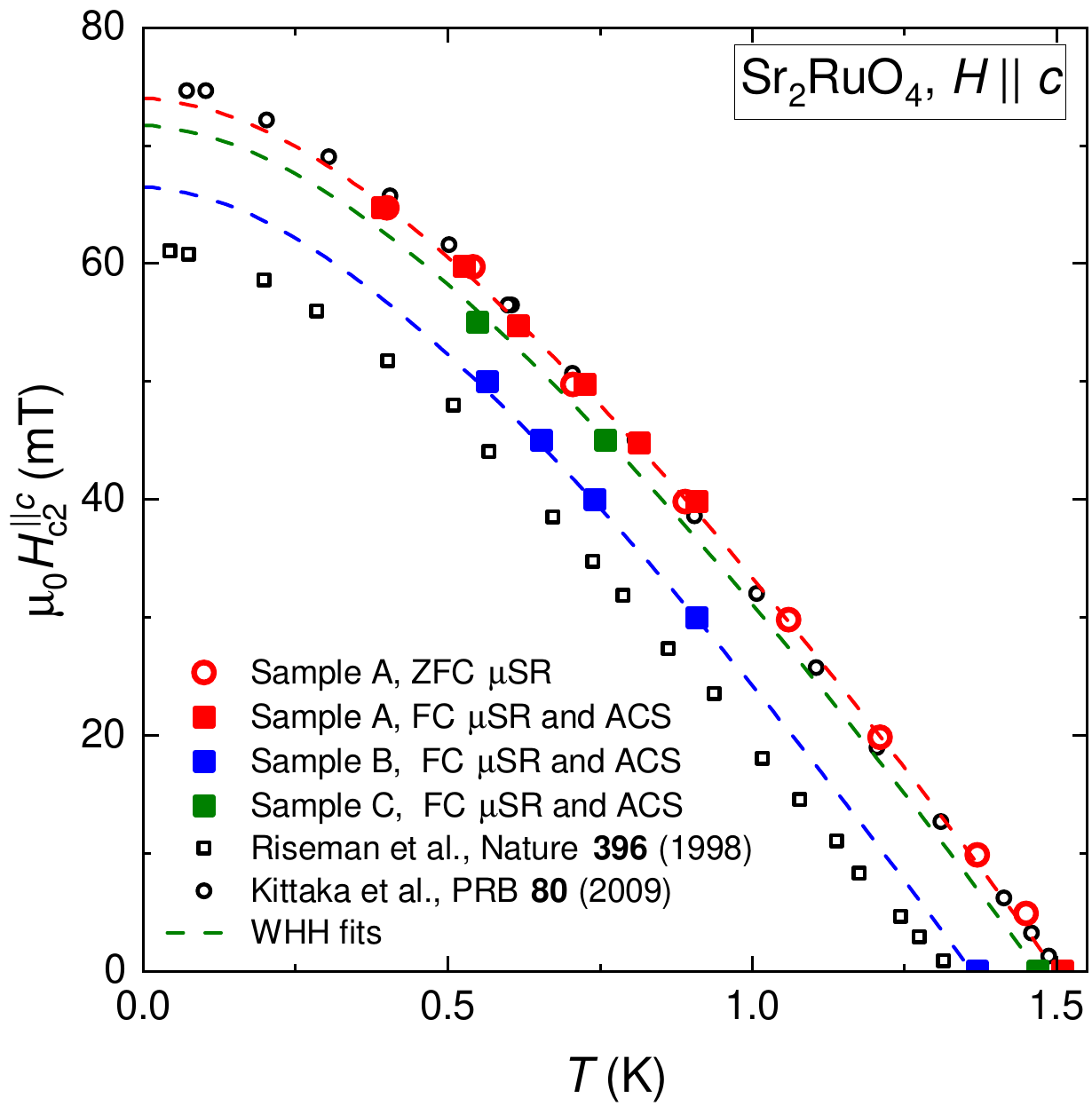}
\caption{\label{fig:hc2}
The temperature dependence of the upper critical field $H_{\rm c2}^{\parallel c}$ in $H\parallel c$ for samples A--C. Literature data from Riseman \textit{et al.} \cite{Riseman1998} and Kittaka \textit{et al.} \cite{Kittaka2009} are shown for comparison. Dashed curves are fits based on the Werthamer-Helfand-Hohenberg (WHH) form \cite{Werthamer1966}.}
\end{figure}

The resulting $H_{\rm c2}^{\parallel c}(T)$ data are summarized in Fig.~\ref{fig:hc2}. For comparison, we also show the most complete literature data sets for \sro\ crystals with $T_{\rm c}=1.28$~K and 1.50~K from Refs.~\onlinecite{Riseman1998,Kittaka2009}. The zero-field transition temperatures of the present samples are either close to the higher-\tc\ reference crystal, as for sample A with \tc=1.51~K, or lie between the two reference data sets, as for samples B and C with \tc=1.48 and 1.40~K, respectively. The agreement with these literature $ H_{\rm c2}^{\parallel c}(T)$ curves, together with the independently determined zero-field transition temperatures from ac susceptibility, shows that the transitions identified by TF-\musr\ correspond to the bulk superconducting transition.

The dashed lines in Fig.~\ref{fig:hc2} are fits based on the Werthamer-Helfand-Hohenberg (WHH) form \cite{Werthamer1966}. In performing these fits, we also take into account the scaling reported in Ref.~\onlinecite{Li2025Preprint}, namely that $H_{\rm c2}^{\parallel c}(T)$ curves for samples with different \tc\ values collapse approximately onto a common curve when normalized by the zero-field superconducting transition temperature. The fits yield $\mu_0\hcTwo^{\parallel c}(0)$ values of 74.0(1), 71.7(1), and 66.5(1)~mT for samples A, B, and C, respectively (see also Table~\ref{tab:sample_parameters}). Thus, for $H\parallel c$, fields exceeding approximately 75~mT suppress bulk superconductivity in the present crystals down to the lowest temperatures. 

\begin{table}[htb]
\caption{\label{tab:sample_parameters}
Summary of the superconducting transition temperatures, $T_{\rm c}$, upper critical fields, $H_{\rm c2}^{\parallel c}(0)$, and characteristic onset temperatures of the exponential relaxation, $T_\Lambda^{\rm ZF}$ and $T_\Lambda^{{\rm TF},80{\rm mT}}$, for samples A--D.}
\begin{ruledtabular}
\begin{tabular}{ccccc}
Sample &
$\tc$ &
$\mu_0\hcTwo^{\parallel c}(0)$ &
$\tlambda^{\rm ZF}$ &
$\tlambda^{{\rm TF,}80{\rm mT}}$ \\
&
(K) &
(mT) &
(K) &
(K) \\
\hline
A & $1.51(1)$ & $74.0(1)$ & $1.14(6)$ & $1.15(7)$ \\
B & $1.48(1)$ & $71.7(1)$ & $1.04(6)$ & $1.06(7)$ \\
C & $1.40(1)$ & $66.5(1)$ & $1.14(6)$ & $1.07(6)$ \\
D & $1.35(2)$ & --        & $1.13(5)$ & --        \\
\end{tabular}
\end{ruledtabular}
\end{table}

\subsection{Comparison of zero-field and 80~mT relaxation}

\begin{figure}[htb]
\centering
\includegraphics[width=0.85\linewidth]{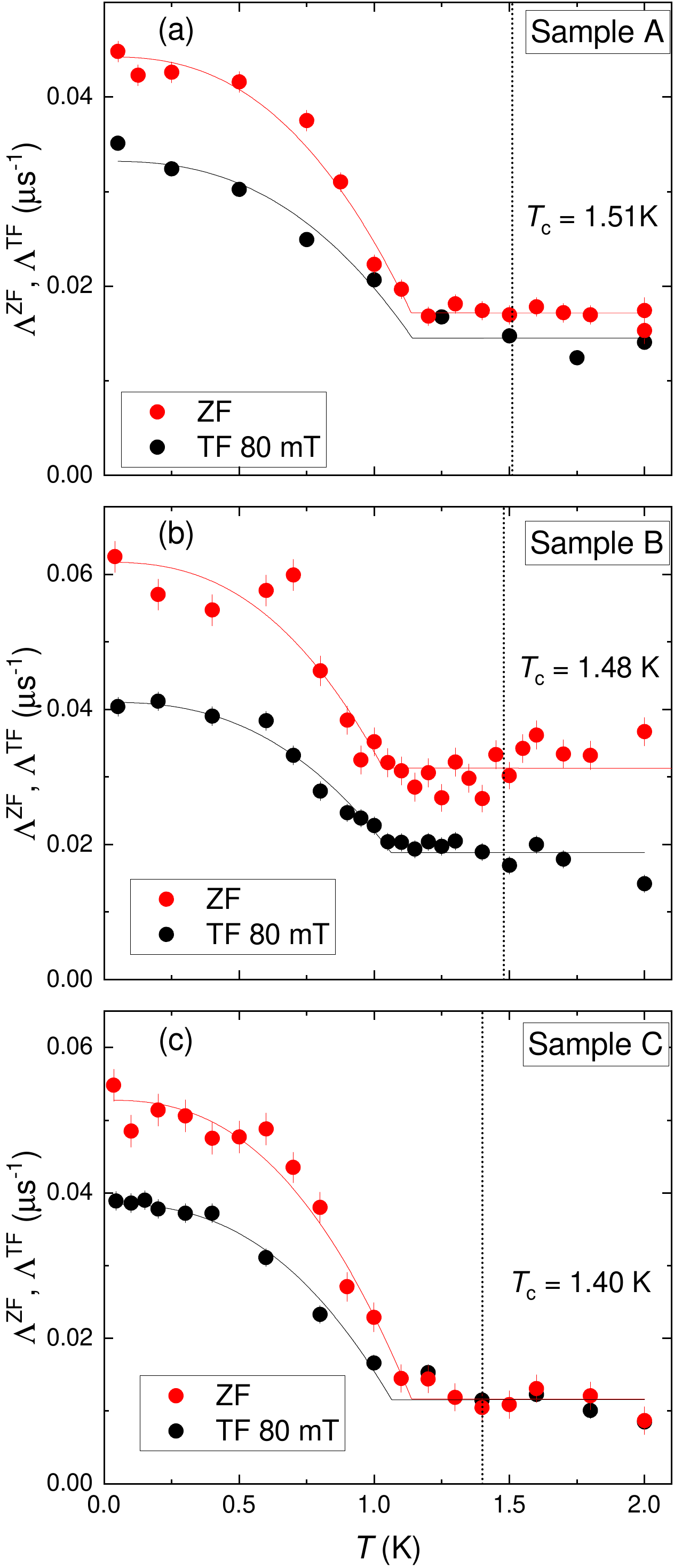}
\caption{\label{fig:zf_tf80}
Temperature dependence of the exponential relaxation rate $\Lambda$ obtained from ZF- and 80~mT TF-\musr\ measurements for (a) sample A, (b) sample B, and (c) sample C. The vertical dotted lines mark the bulk superconducting transition temperatures. Solid curves are fits to Eq.~(\ref{eq:lambda}). }
\end{figure}

The central result of this work is summarized in Fig.~\ref{fig:zf_tf80}, where the exponential relaxation rates measured in ZF and in TF at 80~mT are compared directly for samples A--C. The importance of the 80~mT data is that this field exceeds $\hcTwo^{\parallel c}(0)$ for all three samples, see Table~\ref{tab:sample_parameters} and Fig.~\ref{fig:hc2}. Therefore, the samples are in the normal state, and any relaxation observed at this field cannot originate from the superconducting mixed state or from spontaneous fields associated with a homogeneous superconducting TRSB phase.

Several important points follow from the comparison of ZF and TF 80~mT data:
\begin{itemize}

\item The ZF exponential relaxation rate $\Lambda^{\rm ZF}$ starts to increase only below a temperature \tlambda, which is lower than the bulk superconducting transition temperature \tc. Such behavior is observed for samples A--C in Fig.~\ref{fig:zf_tf80}, and also for sample D, as shown in the Supplementary Information. A possible spectrometer- or cryostat-related origin of this effect was ruled out by repeating the ZF-\musr\ measurements on sample B at the VMS instrument, see the Supplementary Information.

\item A very similar temperature dependence is observed for the exponential relaxation rate measured in TF at 80~mT. The increase of $\Lambda^{\rm TF}$ starts at nearly the same temperature as the increase of $\Lambda^{\rm ZF}$, even though bulk superconductivity is completely suppressed at this field.

\item Fits of Eq.~(\ref{eq:lambda}) to $\Lambda^{\rm ZF}(T)$ and $\Lambda^{\rm TF}(T)$ yield closely similar onset temperatures in the two geometries, see Table~\ref{tab:sample_parameters} and Fig.~\ref{fig:zf_tf80}. Note that to reduce correlations between fitting parameters, the final analysis used a common phenomenological exponent fixed at $n=2.5$, consistent with preliminary fits in which $n$ was allowed to vary independently; see the Supplementary Information.

\item The absolute increase of the exponential relaxation, $\Delta\Lambda$, is larger in ZF than in TF at 80~mT. However, the ratio
    \begin{equation}
    \Delta\Lambda^{\rm ZF}/\Delta\Lambda^{{\rm TF,}80{\rm mT}}\simeq 1.4\text{--}1.5 \nonumber
    \end{equation}
    is approximately the same for all three samples. 

\item Above \tlambda, the exponential relaxation rate remains finite in both ZF and 80~mT TF measurements. In ZF, $\Lambda^{\rm ZF}(T)$ is nearly temperature independent in this range, whereas in TF, $\Lambda^{{\rm TF},80{\rm mT}}(T)$ shows a weak but systematic decrease with increasing temperature above $\tlambda$. The presence of this finite relaxation above the apparent onset temperature indicates that the underlying relaxation channel is already present in the normal state. The increase below \tlambda\ therefore reflects an enhancement or slowing down of a pre-existing channel, rather than the sudden appearance of a completely new contribution.

\item When the extracted \tlambda\ values are plotted as a function of \tc\ in Fig.~\ref{fig:literature}, the present data follow the tendency observed for several other \sro\ samples, namely that \tlambda\ remains separated from \tc.

\end{itemize}

Overall, the close correspondence between the ZF and 80~mT TF relaxation rates shows that a substantial part of the exponential relaxation cannot be assigned uniquely to spontaneous superconducting fields. Instead, the data indicate that the low-temperature increase of $\Lambda$ originates from a normal-state relaxation mechanism that persists when superconductivity is suppressed.

\begin{figure}[htb]
\centering
\includegraphics[width=0.85\linewidth]{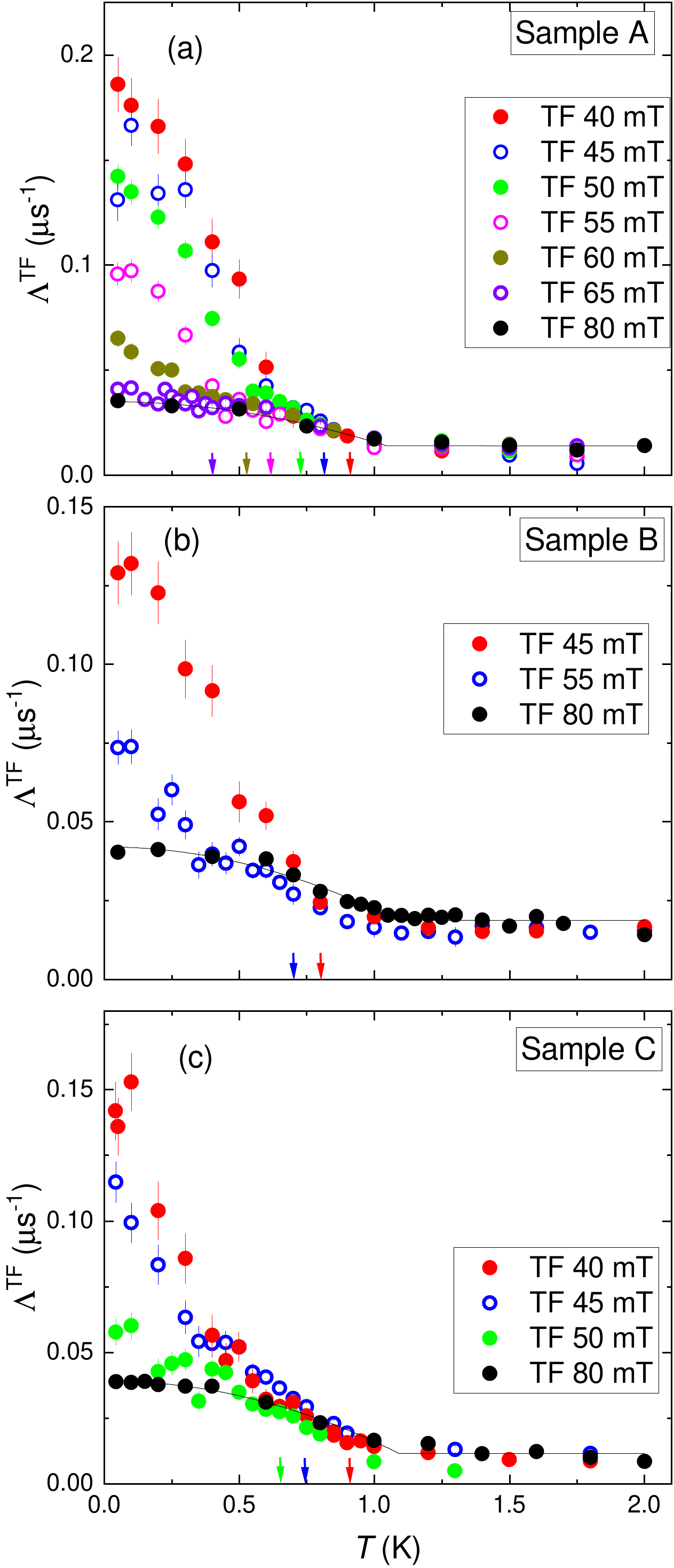}
\caption{\label{fig:tf_lowfield}
Exponential relaxation rate obtained in TF-$\mu$SR experiments below and above $ H_{\rm c2}^{\parallel c}(0)$ for (a) sample A, (b) sample B, and (c) sample C. The large low-temperature relaxation in fields below $ H_{\rm c2}^{\parallel c}(0)$ is attributed primarily to the superconducting mixed state.  The solid lines are fits of Eq.~(\ref{eq:lambda}) to the 80~mT TF data. The arrows indicate the superconducting transition temperature $T_{\rm c}$ determined from the temperature dependences of the Gaussian relaxation rate $\sigma^{\rm TF}$ and the internal field $B_{\rm int}$; see the Supplementary Information.}
\end{figure}

\subsection{Exponential relaxation in the superconducting and normal states}

Figure~\ref{fig:tf_lowfield} shows the temperature dependence of $\Lambda^{\rm TF}(T)$ obtained in TF-\musr\ measurements performed on samples A--C in fields below $ H_{\rm c2}^{\parallel c}(0)$, together with the 80~mT data for comparison. In fields below $ H_{\rm c2}^{\parallel c}(0)$, superconductivity is clearly detected by \musr: on cooling through the field-dependent superconducting transition, the internal field decreases and the Gaussian relaxation rate increases, as expected from the formation of a flux-line lattice (see Fig.~S9 in the Supplementary Information). The corresponding $T_{\rm c}(H)$ values are indicated by colored arrows in Fig.~\ref{fig:tf_lowfield}. The exponential relaxation rate starts to increase already above $\tc(H)$, reflecting the underlying normal-state relaxation channel discussed above. On further cooling through $\tc(H)$, a much stronger additional increase develops in fields below $ H_{\rm c2}^{\parallel c}(0)$. Because this additional contribution is observed only below the upper critical field and follows closely the field-dependent superconducting transition, it should be regarded primarily as a superconducting mixed-state contribution.

The comparison with the 80~mT data is important for separating this mixed-state superconducting contribution from the normal-state relaxation channel discussed above. At 80~mT, bulk superconductivity is suppressed in all three samples, yet a weaker and more gradual low-temperature exponential relaxation remains visible. Therefore, the exponential relaxation measured below $ H_{\rm c2}^{\parallel c}(0)$ should be regarded as a superposition of two terms: a superconducting vortex contribution and an underlying normal-state exponential channel. For each sample, the 80~mT data provide an empirical estimate of this normal-state channel. In the lower-field TF data, the weak underlying relaxation starts to increase at approximately the same temperature as in the 80~mT data, namely near $T_\Lambda^{{\rm TF},80{\rm mT}}$, within the experimental uncertainty.

\begin{figure}[htb]
\centering
\includegraphics[width=0.85\linewidth]{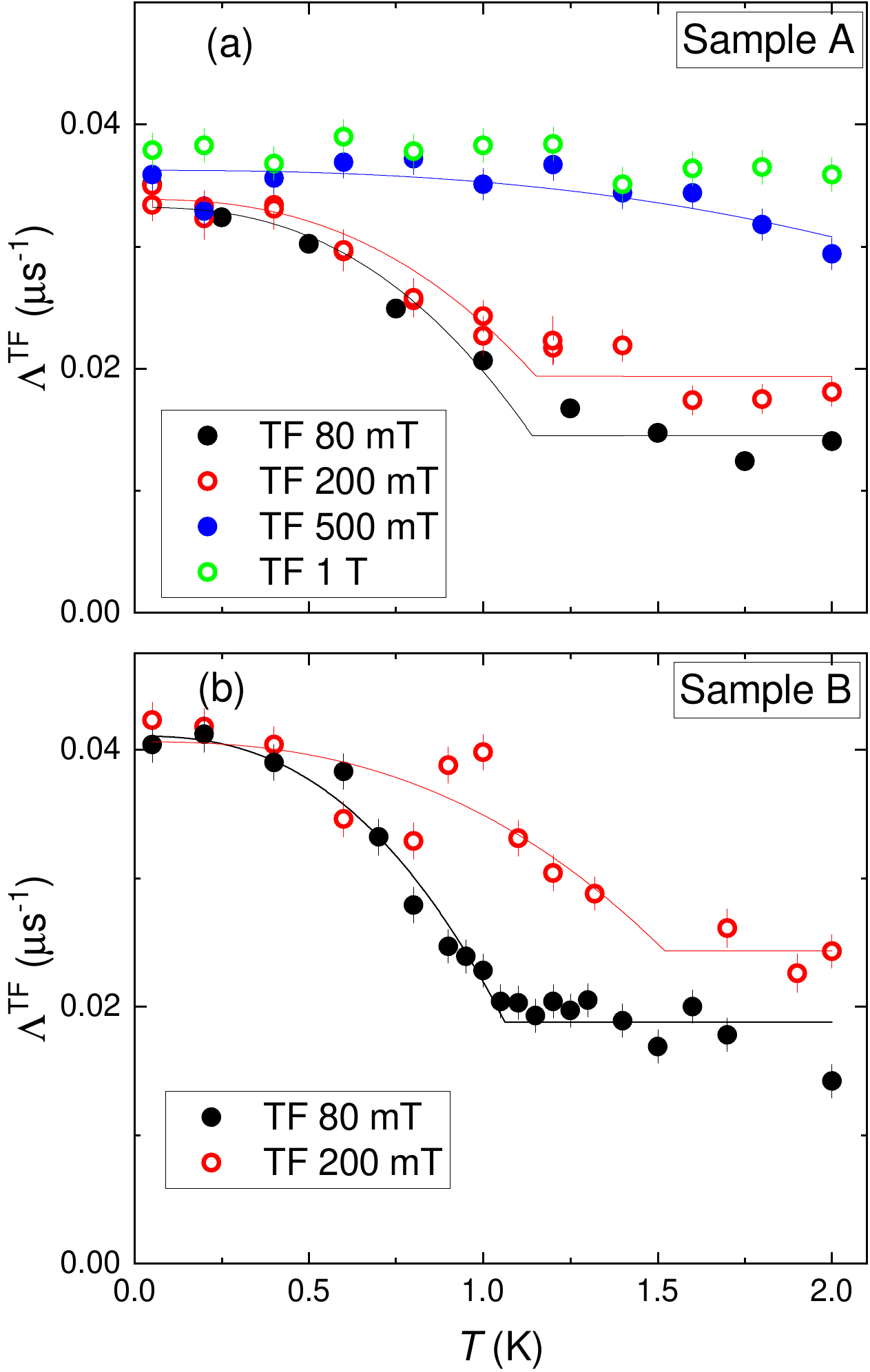}
\caption{\label{fig:tf_highfield}
Temperature dependence of the exponential relaxation rate in transverse-field measurements above $ H_{\rm c2}^{\parallel c}(0)$ for (a) sample A and (b) sample B. The solid lines are fits of Eq.~(\ref{eq:lambda}) to the data.}
\end{figure}

To further characterize this normal-state channel, we measured the temperature dependence of the exponential relaxation in constant transverse fields exceeding $ H_{\rm c2}^{\parallel c}(0)$. The results are summarized in Fig.~\ref{fig:tf_highfield}. For sample A, measurements were performed at 80, 200, 500~mT, and 1~T. The data show that the characteristic temperature scale associated with the change of $\Lambda(T)$ tends to increase with increasing applied field. At 80 and 200~mT, the increase of $\Lambda^{\rm TF}$ on cooling occurs within the measured temperature range. At 500~mT, the onset is shifted to temperatures above about 2~K, so that only the low-temperature part of the relaxation enhancement is observed. At 1~T, $\Lambda^{\rm TF}(T)$ changes only weakly over the accessible temperature window, consistent with the characteristic onset being shifted beyond the measured range. The data for sample B show an analogous comparison between 80 and 200~mT, with the higher-field data exhibiting a broader and higher-temperature evolution of $\Lambda^{\rm TF}(T)$.

\subsection{Longitudinal-field $\mu$SR experiments}

LF-\musr\ measurements were performed to determine whether the internal fields responsible for the exponential relaxation are static/quasistatic or dynamic on the \musr\ time scale. In a conventional LF geometry, a longitudinal field applied parallel to the initial muon-spin polarization decouples the muon spins from a static or quasistatic distribution of local fields, whereas relaxation caused by fast fluctuations is not suppressed in the same way \cite{Hayano1979,Yaouanc2011}.

The most complete LF-\musr\ data set was obtained for sample B, for which the applied longitudinal field was varied from 0.075 to 200~mT. For samples A and C, LF-\musr\ measurements were performed over more limited field intervals, from 5~mT to 1~T for sample A and from 40 to 80~mT for sample C.

\begin{figure}[htb]
\centering
\includegraphics[width=0.85\linewidth]{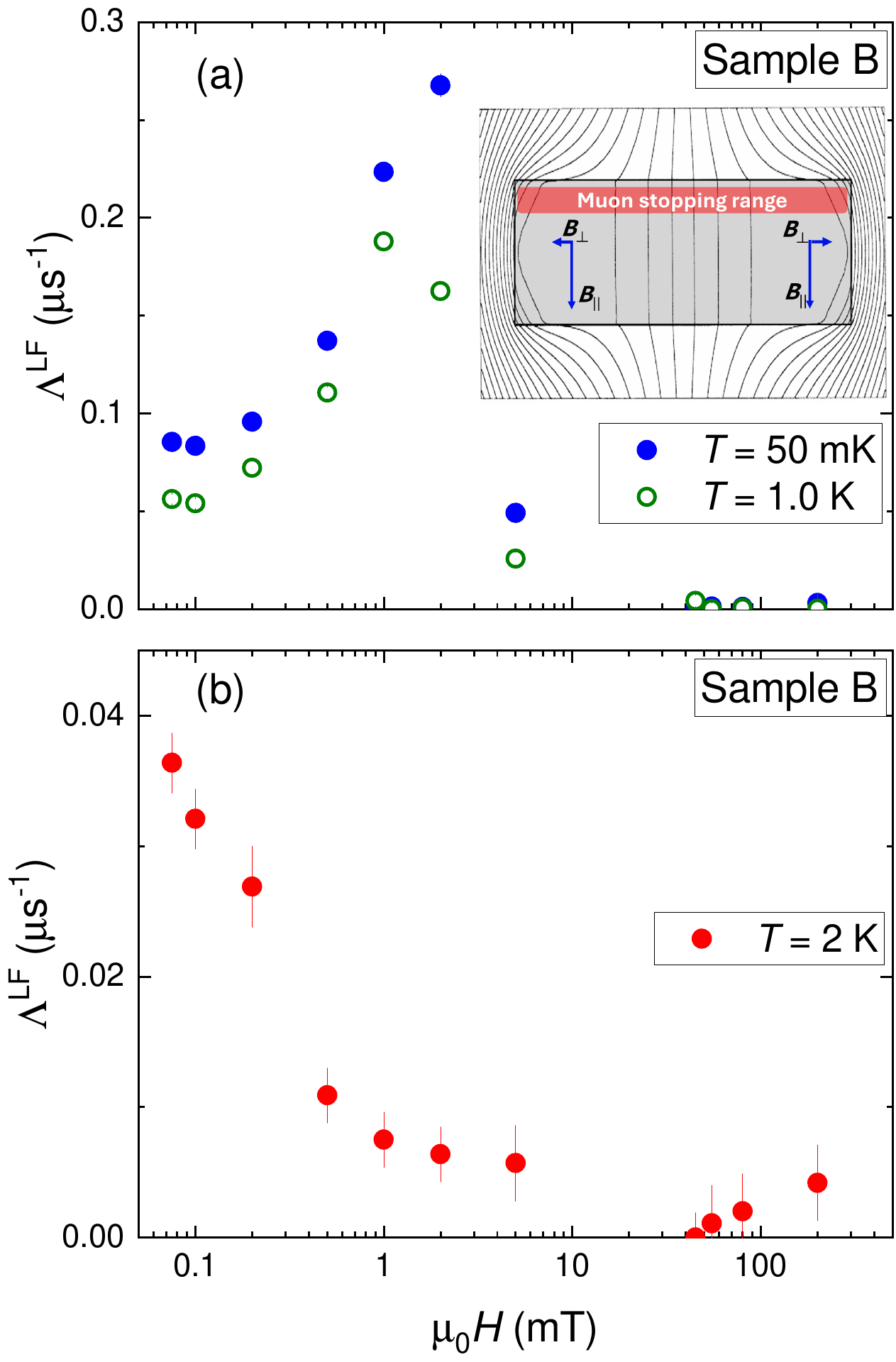}
\caption{\label{fig:lf}
Longitudinal-field dependence of the exponential relaxation rate $\Lambda^{\rm LF}$ for sample B at (a) 50~mK and  1.0~K, and (b) 2.0~K.  The inset in panel (a), adapted from Brandt \cite{Brandt1996}, schematically illustrates the field distribution in a rectangular superconducting sample affected by partial diamagnetism and a geometrical barrier. Field bending near the sample edges can generate local components perpendicular to the nominal LF direction and can therefore contribute to the LF relaxation.}
\end{figure}

The field dependence of $\Lambda^{\rm LF}$ for sample B is shown in Fig.~\ref{fig:lf}. The most direct information is obtained at $T=2$~K, i.e., above the superconducting transition temperature of sample B, \tc=1.48~K. In this normal-state data set, Fig.~\ref{fig:lf}(b), the relaxation is strongly reduced already by longitudinal fields of order 1~mT and is nearly suppressed at higher fields. Such behavior is characteristic of decoupling from a static or quasistatic distribution of local magnetic fields. It disfavors a purely fast-fluctuating origin of the exponential relaxation and instead points to slowly fluctuating or static dilute magnetic fields, or to a temperature-dependent change of the static field distribution sampled by the muons.

The LF data below \tc, shown in Fig.~\ref{fig:lf}(a), require additional care. In a nonsuperconducting sample, a longitudinal field applied parallel to the initial muon-spin polarization decouples static transverse local fields in a relatively direct way. In the superconducting state, however, the internal magnetic field in a real sample is generally not uniform. Partial diamagnetism, demagnetization effects, vortex penetration, and geometrical barriers can all distort the field distribution. Among these effects, the geometrical barrier is expected to be particularly relevant for the present platelet-like crystals. Such a barrier is absent only for ideal ellipsoidal samples, for which the demagnetizing field is homogeneous. In nonellipsoidal samples, by contrast, it affects flux entry and can produce a strongly inhomogeneous spatial distribution of vortices and internal fields \cite{Zeldov1994GB,Brandt1996,Brandt1999,Brandt2000,BrandtMikitikZeldov2013}.

The inset of Fig.~\ref{fig:lf}(a), adapted from Brandt \cite{Brandt1996}, illustrates this effect for a rectangular superconducting sample. The field lines are distorted by the geometrical barrier: far from the sample they follow the applied field direction, whereas near the sample edges they bend strongly. Consequently, muons stopping near regions with strong field curvature can experience local fields that are not parallel to the applied LF field. This effect should be less pronounced in large and thin single crystals, where the relative weight of the edge regions is reduced, but it remains an important complication for finite rectangular samples.

To summarize, the normal-state LF-\musr\ measurements provide strong evidence that the exponential relaxation originates from static or quasistatic local fields. In the superconducting state, the LF response is compatible with this conclusion, but a quantitative separation of microscopic magnetism from superconductivity-induced field inhomogeneity requires caution because geometrical barriers and vortex-field distortions can themselves generate transverse local-field components in a nominal LF experiment.

\section{Discussion}

In this section, we first summarize the results of the
muon stopping-site calculations and estimate the ZF relaxation expected from
the nuclear moments of naturally abundant O, Sr and Ru isotopes. We then
discuss possible mechanisms that may account for the observed temperature
and field evolution of the low-temperature exponential relaxation. Finally,
we consider the consequences of our experimental findings for the
interpretation of time-reversal-symmetry breaking in \sro.

\subsection{Muon stopping sites and nuclear contribution to the ZF relaxation}


\subsubsection{Muon stopping sites and hopping}

To determine the muon stopping sites, we carried out
density-functional theory (DFT) calculations using the plane-wave basis-set
electronic-structure code CASTEP~\cite{Clark2005}. Muon-site calculations
were performed using the MuFinder program~\cite{Huddart2022}. We confirm
the results proposed in Ref.~\cite{Huddart2021}, namely that the muon stops
approximately 0.97~\AA{} from an apical oxygen of a RuO$_6$ octahedron.

\begin{figure}[hb]
\includegraphics[width=1.0\linewidth]{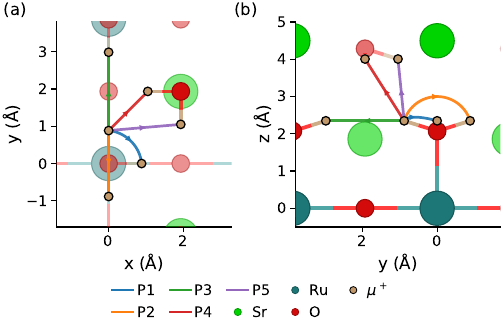}
\caption{
Candidate muon hopping paths in Sr$_2$RuO$_4$, viewed (a) along the $c$ axis
and (b) along the $a$ axis. The paths connect symmetry-equivalent copies of
the lowest-energy muon site. Atomic layers further away from the viewer are
shown with increasing transparency.
}
\label{fig:muon-hopping}
\end{figure}

Among five possible hopping paths connecting symmetry-equivalent muon sites
separated by less than 3.0~\AA{}, the shortest path, with a site-to-site
distance of 1.25~\AA{}, has the lowest barrier, $V_0 \simeq 70$~meV.
This barrier becomes effectively zero once the zero-point energies of the
muon at the stopping site and at the top of the barrier are taken into
account. By contrast, paths connecting sites associated with different
apical oxygen atoms have an effective barrier of approximately 0.37~eV,
making long-range muon diffusion negligible at the temperatures of our
measurements. The muon should therefore be considered as delocalized over
four symmetry-equivalent sites around an apical oxygen; see
Fig.~\ref{fig:muon-hopping}. Further details of the calculations, including
the candidate muon hopping paths, transition states, and
zero-point-energy-corrected barriers, are given in the Supplementary Information.

\subsubsection{Nuclear contribution to the zero-field relaxation}

The exponential relaxation observed in ZF-\musr,
including the finite relaxation already present above \tc, is striking and
reproducible between different samples. Its suppression by modest
longitudinal fields indicates that the underlying local fields are
predominantly static or quasistatic. One possible source of static local
fields is the nuclear moments.

Oxygen has no nuclear moment, apart from the 0.04\%
natural abundance of $^{17}$O, whereas $^{87}$Sr, $^{99}$Ru, and
$^{101}$Ru have finite nuclear moments and natural abundances of
approximately 7\%, 13\%, and 17\%, respectively. The remaining isotopes have
$I=0$. This corresponds to a relatively dilute background of nuclear spins.
We simulated the relaxation expected from these nuclear moments using a
Monte Carlo method. The result, shown in Fig.~S12 of the Supplementary Information, is much closer to a Gaussian-like relaxation than to the
experimentally observed exponential form. In addition, its magnitude is too
small to account for the measured relaxation. We therefore conclude that
static nuclear moments alone cannot explain the observed exponential character of the ZF-\musr\
relaxation.

\subsection{Possible sources of the low-temperature relaxation}
\label{sec:possible_mechanisms}

Our data impose several constraints on the possible
microscopic origin of the relaxation. The relevant mechanism must already be
present in the normal state, must produce similar temperature dependences in
ZF-\musr\ and in TF-\musr\ at 80~mT, must lead to a smaller relaxation
amplitude in TF than in ZF, must be nearly static at low temperatures, and
must account for the exponential rather than Gaussian form of the ZF
relaxation. We therefore consider below several possible sources of the
low-temperature relaxation in light of these constraints.

The scenarios considered below include dilute or
defect-related magnetism associated with Sr$_2$RuO$_4$ itself,
surface-related altermagnetism, macroscopic or microscopic impurity phases,
and effects associated with the implanted muon, such as trapping, hopping,
or diffusion.

\subsubsection{Dilute magnetic moments or defect-related magnetism in Sr$_2$RuO$_4$}
\label{sec:Dilute_magnetism}

Since the simulated nuclear contribution is both too weak
and closer to a Gaussian-like relaxation, we next consider a dilute
distribution of electronic moments. For a static dilute distribution of
randomly located moments, the relaxation may approach a Lorentzian
Kubo--Toyabe form~\cite{Fiory1984,Gist1984},
$G(t)=\frac{1}{3}+\frac{2}{3}(1-at)\exp(-at)$, which is approximately
exponential at early times, $G(t)\simeq \exp(-\Lambda t)$, with
$\Lambda=4a/3$; see the Supplementary Information for details. Taking $\Lambda\simeq 0.04~\mu{\rm s}^{-1}$ and
moments of order $1~\mu_{\rm B}$ gives an estimated magnetic-site
concentration $c\simeq 800$~ppm. Such dilute electronic moments could be
associated with Ru-site vacancies, antisite defects, oxygen defects, or
local strain fields that polarize nearby Ru orbitals. In this picture, a
nominally nonmagnetic structural defect could generate a spatially extended
Ru spin polarization.

The LF-\musr\ data provide the most direct constraint on this interpretation.
As shown in Fig.~\ref{fig:lf}, the relaxation is rapidly suppressed by small
longitudinal fields at 2~K. A weak longitudinal field can decouple the muon
spins from a static or slowly fluctuating distribution of dilute local fields
\cite{Hayano1979,Yaouanc2011}. The observed LF decoupling therefore
indicates that, once the low-temperature relaxation has developed, the fields
sensed by the muons are predominantly static or quasistatic on the \musr\
time scale.

The comparison between the ZF and TF relaxation amplitudes gives an
additional consistency check. In ZF, the muon polarization is relaxed by the
full distribution of randomly oriented local fields. In a transverse applied
field, the external field defines the dominant precession axis and partially
averages or decouples the effect of the local-field components perpendicular
to it. For a static or quasistatic dilute field distribution, the TF
relaxation is therefore expected to be reduced compared with the ZF
relaxation, while retaining a similar temperature dependence if both
measurements probe the same underlying magnetic source \cite{Yaouanc2011}.
The observation that $\Delta\Lambda^{\rm TF,80mT}$ is smaller than
$\Delta\Lambda^{\rm ZF}$, but both $\Lambda^{\rm ZF}(T)$ and
$\Lambda^{\rm TF,80mT}(T)$ follow a comparable temperature dependence, is
consistent with this expectation.

The field dependence of the normal-state TF relaxation provides further
information on the origin of the relevant relaxation channel.
Figure~\ref{fig:tf_lowfield} indicates that $\Lambda^{\rm TF}(T)$ starts to
increase already above $T_{\rm c}(H)$, as part of the same relaxation channel
that is also present at 80~mT. As shown in Fig.~\ref{fig:tf_highfield}, the
characteristic temperature scale associated with the increase of
$\Lambda^{\rm TF}(T)$ appears to shift to higher temperatures with increasing
applied field. Since the high-field measurements were performed above
$H_{\rm c2}^{\parallel c}(0)$, this shift is not related to the superconducting
mixed state. A possible interpretation is that the applied field polarizes
dilute moments, magnetic clusters, or defect-related magnetic regions, thereby
modifying the quasistatic local-field distribution sensed by the muons. In
this sense, the field dependence may reflect a field-tuned crossover of a
normal-state magnetic background rather than a superconducting transition.
Thus, within the scenarios discussed in this subsection, the field dependence
of $T_\Lambda$ should not be interpreted as evidence for a purely
fast-fluctuating Redfield-type process, but rather as a possible field tuning
of a static or quasistatic local-field distribution.

This explanation is related, but not identical, to defect-induced TRSB models
\cite{Willa2021}. In such models, a primarily single-component
superconducting state can acquire local complex components near strain fields
or dislocations, generating spontaneous fields in an inhomogeneous manner.
The persistence of the 80~mT response when bulk superconductivity is
suppressed shows that such a superconducting mechanism cannot account for
the entire relaxation. Nevertheless, local defects remain a plausible common
ingredient in both superconducting and non-superconducting interpretations.

\subsubsection{Surface-related altermagnetism}
\label{sec:surface_altermagnetism}

Recent theoretical works have proposed altermagnetism as a possible origin
of weak time-reversal-symmetry-breaking magnetism in Sr$_2$RuO$_4$,
especially in connection with local symmetry lowering caused by surface
reconstruction and rotations of the RuO$_6$ octahedra
\cite{Autieri_PRB_2025,Ramires_arXiv_2025}. Such a distortion may
stabilize a compensated magnetic state with no, or only very small, net
magnetization, while still producing weak static or quasistatic local fields
detectable by $\mu$SR.

These theoretical proposals are closely connected to the earlier low-energy
(LE) $\mu$SR observation of weak static surface magnetism in
Sr$_2$RuO$_4$~\cite{Fittipaldi_NatCommun_2021}. This is relevant here
because the LE-$\mu$SR response shows several qualitative similarities to
the additional exponential relaxation observed in our measurements: it is
seen in the normal state, appears in TF-$\mu$SR, and LF-$\mu$SR indicates
that the associated local fields are static or quasistatic.

A further important point is the depth dependence of the
LE-$\mu$SR signal. The strongest increase of the relaxation rate was
observed for muons stopping within approximately 10--20~nm from the surface,
with an onset temperature of $\sim 50$--75~K. At larger implantation depths,
such as about 70~nm, the effect becomes weaker and appears only below
approximately 25~K~\cite{Fittipaldi_NatCommun_2021}. This suggests that the
characteristic temperature scale detected by $\mu$SR may depend on the muon
stopping depth. The so-called ``surface'' muons used in the present
experiment have an energy of $\simeq 4$~MeV and therefore stop much deeper
in the sample, at typical depths of order 0.1--0.2~mm (see, e.g., Fig.~2 in
Ref.~\onlinecite{Khasanov_JAP_2022}).

A connection between these observations remains speculative. The present
data do not provide direct evidence for altermagnetism, and the large
difference between LE-$\mu$SR and conventional $\mu$SR stopping depths
excludes a simple one-to-one correspondence. Such a connection would require
that surface-related magnetism, or a related defect- or strain-induced
magnetic state, gives rise to weak static local fields also within the bulk,
for example in the vicinity of stacking faults, dislocations, structural
domains, or other locally strained regions. In this context, altermagnetism
may provide a possible microscopic route to the weak static magnetic
background observed in Sr$_2$RuO$_4$.

\subsubsection{Role of macroscopic and microscopic impurity phases}
\label{sec:impurity-phases}

A possible extrinsic source of local magnetic fields in \sro\ is the
presence of secondary phases from the Sr--Ru--O phase diagram. Depending
on small deviations from the ideal growth conditions, crystals may contain
traces of Ru lamellae, SrRuO$_3$ (the 113 phase), Sr$_3$Ru$_2$O$_7$ (the
327 phase), or other higher-order Ruddlesden--Popper ruthenates. These
phases can occur either as macroscopic inclusions or, more subtly, as
microscopic intergrowths or stacking-fault defects within the
Sr$_2$RuO$_4$ matrix.

Ru-metal inclusions represent a well-known example of such secondary
phases. They are known to enhance the local superconducting transition
temperature to $T_{\rm c}\simeq 3$~K and the upper critical field to
$H_{\rm c2}^{\parallel c}(0)\simeq 1$~T~\cite{Maeno_PRL_1998}. The
ac-susceptibility results shown in Fig.~S3 of the Supplementary Information
demonstrate that the samples used in this study contain only a negligible
amount of such Ru inclusions. Ru-metal inclusions are therefore unlikely to
be responsible for the observed low-temperature exponential \musr\
relaxation.

We next consider ruthenate impurity phases related to the 113 and 327
structures. The 113 phase is ferromagnetic, while the 327 phase exhibits
strongly enhanced paramagnetic susceptibility. Their presence can therefore
be tested by dc-magnetization measurements, providing an independent way to
assess whether macroscopic SrRuO$_3$ or Sr$_3$Ru$_2$O$_7$ impurity phases
could be responsible for the observed exponential \musr\ relaxation.

In addition to macroscopic impurity phases, 113- or 327-like structural
fragments may also occur as microscopic intergrowths or stacking-fault
defects within the Sr$_2$RuO$_4$ matrix. In this case, local deviations
from the ideal 214 stacking sequence can produce fragments resembling the
113, 327, or other related Ruddlesden--Popper phases. The volume fraction
of such microscopic intergrowths is expected to be much smaller than that
of resolvable macroscopic impurity phases, and they may therefore be
difficult to detect by standard bulk characterization. Nevertheless, if
they are magnetic or strongly polarizable, they could contribute to an
inhomogeneous static or quasistatic local-field distribution sensed by the
muons.

To test the role of macroscopic 113 and 327 impurity phases, we performed
dc-susceptibility measurements on pieces cut from the same growth rods as
the \musr\ samples. The results, shown in Fig.~S4 of the Supplementary Information, reveal sample-dependent low-level impurity contributions. In
particular, a magnetic onset near 160~K is consistent with traces of the
113 phase, while a feature around 20~K is characteristic of the 327 phase.

The relative magnitude of these impurity signatures differs substantially
between the samples. For example, sample A exhibits comparatively strong
contributions associated with the 113 and 327 phases, corresponding to
estimated impurity levels of approximately 600~ppm and 1.3\%, respectively.
By contrast, sample C shows the weakest signatures of both 113 and 327
impurity phases. This variation does not correlate with the \musr\ results:
the low-temperature increase of the exponential relaxation rate $\Lambda$
is observed in all samples and displays a broadly similar temperature
dependence despite the substantially different impurity signatures inferred
from the magnetization measurements. Moreover, the relaxation enhancement
$\Delta\Lambda$ does not scale with the magnitude of the macroscopic
impurity signatures. In particular, although sample C shows only weak 113
and 327 contributions in dc magnetization, its low-temperature increase of
$\Lambda$ is comparable to, or even larger than, that observed in samples
with substantially stronger impurity signatures, such as samples A and B.
We therefore conclude that macroscopic SrRuO$_3$ or Sr$_3$Ru$_2$O$_7$
impurity phases are unlikely to be the primary origin of the observed
exponential \musr\ relaxation.

This conclusion does not completely exclude a contribution from microscopic
intergrowths, stacking faults, or other local structural defects that are
not resolved as macroscopic impurity phases. However, such defects would
represent a dilute, spatially inhomogeneous source of static or quasistatic
local fields, rather than a bulk impurity phase. They therefore fall within
the broader class of defect-related magnetic backgrounds discussed above.

\subsubsection{Muon-induced effects}
\label{sec:muon-induced_fields}

Muon trapping, changes of stopping configuration, or local hopping among
muon stopping sites can also produce a temperature-dependent exponential
relaxation
\cite{Yaouanc2011,Hartmann1989MuonDiffusionAl,
Kadono1991MuonDiffusionAl,Pfiz1991MuonDiffusionAl,
Karlsson1995MuonDiffusionAl,MatsumotoOhashi1993MuonDiffusionI,
OhashiMatsumoto1993MuonDiffusionII,Blundell2025,
BlundellLancaster2023}.
The calculations discussed above indicate that long-range diffusion between
sites associated with different apical oxygen atoms is negligible at the
temperatures of our measurements. Local hopping or delocalization among the
four symmetry-equivalent positions around a given apical oxygen, however,
may still modify the local-field distribution sampled by the muon. Upon
cooling, changes in the local muon state or trapping near defects could
similarly lead to an increased relaxation rate.

This mechanism is relevant because the relaxation channel is already
present above \tlambda\ and becomes enhanced on cooling, rather than
appearing abruptly at \tc. The smooth onset, the presence of the effect in
the normal state, and the LF decoupling are compatible with this class of
explanations, provided that the low-temperature muon state samples a
predominantly static or quasistatic local-field distribution. In this case,
muon trapping or a change of site would not itself constitute a fast
dynamic relaxation mechanism, but would instead modify the static local
fields sampled by the muon. Hopping or delocalization among
symmetry-equivalent sites near the apical oxygen could similarly modify the
coupling of the muon to local nuclear moments, electric-field gradients, or
weak electronic moments. In addition, motion of the muon around the charged
oxygen could, in principle, generate an effective magnetic field at the muon
site and modify its response to an applied field. Whether such motion can
remain sufficiently slow or quasistatic at low temperature, and whether it
can account for the observed field dependence, remain open and highly
speculative questions.

This scenario should be tested further by extending the temperature range
above 2~K and by comparing samples with different defect concentrations, for
example after controlled electron or ion irradiation. If muon trapping,
changes in the local stopping configuration, or local hopping are involved,
the characteristic temperature scale of the relaxation should depend on the
local defect environment and possibly on the muon stopping configuration.

\subsection{Consequences for TRSB in \texorpdfstring{\sro}{Sr2RuO4}}

Our ZF-\musr\ data are fully consistent with the previously published
results on \sro. They confirm that the low-temperature increase of the ZF
exponential relaxation rate, appearing at or below \tc, is a robust and
reproducible experimental observation. This anomaly has been reported over
nearly three decades, on different samples and using different instruments
and facilities
\cite{Luke1998,Luke2000,Higemoto2014,Grinenko2021NatCommun,
Grinenko2021NatPhys,Grinenko2023PRB,Khasanov2023PRL,Li2025Preprint}.
The new aspect of the present study is that a closely related exponential
relaxation channel is also detected when bulk superconductivity is
suppressed by an applied field above $ H_{\rm c2}^{\parallel c}(0)$.
This demonstrates that the total relaxation contains a normal-state
contribution, which must be accounted for before assigning the ZF
relaxation increase uniquely to superconducting TRSB.

An earlier hydrostatic-pressure study found that $T_\Lambda$ and \tc\ are
suppressed together~\cite{Grinenko2021NatCommun}, which naturally suggested
a connection between the additional relaxation and superconductivity. This
observation remains important, but it does not by itself establish that the
relaxation is caused directly by a homogeneous TRSB superconducting order
parameter. Experiments under uniaxial stress also indicate a slight suppression
of $T_\Lambda$, at least initially
\cite{Grinenko2021NatPhys,Grinenko2023PRB}. If pressure generally reduces
the characteristic temperature scale of the relaxation channel, the
apparent tracking of $T_\Lambda$ and \tc\ under hydrostatic pressure may
reflect two pressure-dependent quantities that vary in parallel.
Alternatively, the weak static magnetic background may be coupled to the
same low-energy electronic, magnetic, or lattice degrees of freedom that
are affected by pressure and are relevant for superconductivity. This
interpretation remains speculative, but it removes an apparent
contradiction between the pressure results and the present high-field
TF-\musr\ data.

This broader interpretation is also consistent with thermodynamic
studies under uniaxial pressure and strain, which have found no evidence
for an additional bulk transition below $T_{\rm c}$ or for the transition
splitting expected for a homogeneous two-component TRSB superconducting
state~\cite{Li2021PNAS,Li_Nature_2022,Jerzembeck2024}.
Such an absence is difficult to reconcile with a homogeneous two-component
superconducting state in which a separate TRSB transition occurs below \tc,
since such a transition would generally be expected to produce a
thermodynamic anomaly.
Our results offer a possible explanation for this discrepancy: the
low-temperature \musr\ relaxation previously attributed to spontaneous
superconducting fields may be due, at least in part, to a process that is
also present in the normal state. In this case, the onset temperature
\tlambda\ extracted from \musr\ need not correspond to a bulk
superconducting TRSB transition.

The TF data below $ H_{\rm c2}^{\parallel c}(0)$ further support this
interpretation. In the superconducting mixed state, the exponential
relaxation contains both the normal-state channel and an additional
flux-line-lattice-related contribution. In ZF, there is no vortex lattice,
but a related normal-state exponential relaxation channel may still be
present and can mimic an onset of spontaneous internal fields.
Consequently, any claim of superconducting TRSB from \musr\ in \sro\ should
demonstrate that the enhanced relaxation is not caused by a normal-state
contribution, for example by determining the relaxation channel after
superconductivity is suppressed.

\section{Summary}

We studied four \sro\ single crystals from different growth batches,
focusing on the low-temperature increase of the exponential relaxation rate
$\Lambda$, which in previous ZF-\musr\ studies was assigned to the onset of
TRSB superconductivity. Three crystals with superconducting transition
temperatures \tc=1.51, 1.48, and 1.40~K were studied using ZF-, TF-, and
LF-\musr. In addition, ZF-\musr\ data for a fourth crystal with
\tc=1.35~K are discussed in the Supplementary Information. In all four ZF data
sets, $\Lambda$ increases below a characteristic temperature \tlambda\ that
is separated from the bulk superconducting transition temperature. This
behavior follows the trend observed in several other ZF-\musr\ studies of
unstrained \sro, where \tlambda\ is not universally locked to \tc.

DFT calculations confirm that the muon stopping site is
located near an apical oxygen of a RuO$_6$ octahedron, with possible
delocalization over four symmetry-equivalent positions. The relaxation
expected from the naturally abundant nuclear moments of Sr and Ru was
estimated by Monte Carlo simulations. The calculated nuclear contribution is
closer to a Gaussian-like relaxation and is too weak to account for the
experimentally observed exponential relaxation, indicating that static
nuclear moments alone cannot explain the data.

A central result is that a quantitatively similar relaxation channel is
observed in TF-\musr\ at 80~mT. Since this field exceeds
$H_{\rm c2}^{\parallel c}(0)$ for the present
crystals, the samples are in the normal state and the relaxation cannot
originate from the superconducting state itself. The temperature dependence
and onset scale are close to those observed in ZF, while the amplitude of
the relaxation enhancement is smaller in TF than in ZF. This reduction is
consistent with expectations for a static or quasistatic dilute local-field
distribution in an applied transverse field.

Measurements below $H_{\rm c2}^{\parallel c}(0)$
reveal an additional superconducting mixed-state contribution potentially
associated with the flux-line lattice. Thus, the TF relaxation below
$H_{\rm c2}^{\parallel c}(0)$ should be viewed as a
superposition of the normal-state exponential channel and a superconducting
contribution. Measurements above
$H_{\rm c2}^{\parallel c}(0)$ show that the
characteristic temperature scale of the normal-state relaxation is field
dependent, further excluding a simple temperature-independent nuclear
dipolar background. In particular, the apparent \tlambda\ scale tends to
increase with increasing transverse field above
$H_{\rm c2}^{\parallel c}(0)$, suggesting that this
normal-state relaxation channel is field tunable. This behavior may reflect
field-induced polarization or gradual freezing of dilute defect-related
moments, rather than a temperature-independent background.

LF-\musr\ measurements show that the relaxation is strongly suppressed by
small longitudinal fields, indicating that the underlying local fields are
static or quasistatic on the \musr\ time scale. In the superconducting
state, LF measurements must be interpreted with caution because partial
diamagnetism, vortex penetration, and geometrical barriers in finite
rectangular crystals can generate nonuniform internal fields and local
components perpendicular to the nominal LF direction.

Taken together, the data show that the exponential relaxation detected in
ZF-$\mu$SR experiments on \sro\ cannot be assigned uniquely to spontaneous
fields of superconducting origin. The most natural explanations are dilute
magnetic moments or defect-related magnetism.
The muon stopping-site results and the nuclear-spin
simulation support this conclusion by showing that the conventional static
nuclear contribution is insufficient, while the observed exponential form is
more naturally compatible with dilute or spatially inhomogeneous sources of
local fields. A gradual freezing or slowing down of dilute defect-related
moments may contribute to the onset of the relaxation, but the LF data
indicate that the low-temperature state is predominantly static or
quasistatic.

\section*{Acknowledgments}
This work is based on experiments performed at the Swiss Muon Source SµS, Paul Scherrer Institute, Villigen, Switzerland. The authors acknowledge P. Dalmas de R\'eotier, M. Sigrist, J. Sonier, and A. Mackenzie for helpful discussions and A. Ikeda and H. Matsuki for their contributions in characterizing some of the samples.
The work of Y.M. was supported by JP22H01168 and JP26K07012. T.J. acknowledges support
as JSPS International Research Fellow (PE24047).
The work of N.K. was supported by JSPS KAKENHI (No. JP21H01033, No. JP22K19093, and JP24K01461). B.M.H. and S.J.B. acknowledge support from UK Research and Innovation (UKRI) under the UK Government's Horizon Europe funding guarantee [Grant No. EP/X025861/1].

\section*{Author contributions}
R.K. and Y.M. conceived and coordinated the study and conducted the $\mu$SR
experiments. R.K. analyzed the data and wrote the manuscript with input from all
authors. T.J.H., J.A.K., T.S., and H.L. contributed to the $\mu$SR
experiments. J.S.B., T.J., N.K., and Y.M. contributed to sample preparation
and characterization. B.M.H. and S.J.B. performed the muon-site and related
computational calculations and analyzed the results. 
All authors approved
the final version.

\section*{Ethics declaration}
Not applicable.

\section*{Data availability}
The data supporting the findings of this study are available within the
Article and its Supplementary Information. Additional data, including the
underlying experimental data, are available from the corresponding author
upon reasonable request.

\section*{Competing interests}
The authors declare no competing interests.

\end{document}